\documentstyle[preprint,psfig,aps,floats]{revtex}
\tightenlines
\begin{document}
\lefthyphenmin=2
\righthyphenmin=3

\title{Measurement of the Shape of the Transverse Momentum Distribution of $W$
       Bosons Produced in {\mbox{$p\bar p$}}\ Collisions at 
       {\mbox{$\sqrt{s}$ =\ 1.8\ TeV}} }

%
\author{                                                                      
B.~Abbott,$^{31}$                                                             
M.~Abolins,$^{27}$                                                            
B.S.~Acharya,$^{46}$                                                          
I.~Adam,$^{12}$                                                               
D.L.~Adams,$^{40}$                                                            
M.~Adams,$^{17}$                                                              
S.~Ahn,$^{14}$                                                                
H.~Aihara,$^{23}$                                                             
G.A.~Alves,$^{10}$                                                            
N.~Amos,$^{26}$
% ...
\\                                                                            
\vskip 0.50cm                                                                 
\centerline{(D\O\ Collaboration)}                                             
\vskip 0.50cm                                                                 
}                                                                             
\address{                                                                     
\centerline{$^{1}$Universidad de los Andes, Bogot\'{a}, Colombia}             
\centerline{$^{2}$University of Arizona, Tucson, Arizona 85721}               
\centerline{$^{3}$Boston University, Boston, Massachusetts 02215}             
\centerline{$^{4}$Brookhaven National Laboratory, Upton, New York 11973}      
\centerline{$^{5}$Brown University, Providence, Rhode Island 02912}           
\centerline{$^{6}$Universidad de Buenos Aires, Buenos Aires, Argentina}       
\centerline{$^{7}$University of California, Davis, California 95616}          
\centerline{$^{8}$University of California, Irvine, California 92697}         
\centerline{$^{9}$University of California, Riverside, California 92521}      
\centerline{$^{10}$LAFEX, Centro Brasileiro de Pesquisas F{\'\i}sicas,        
                  Rio de Janeiro, Brazil}                                     
\centerline{$^{11}$CINVESTAV, Mexico City, Mexico}                            
\centerline{$^{12}$Columbia University, New York, New York 10027}             
\centerline{$^{13}$Delhi University, Delhi, India 110007}                     
\centerline{$^{14}$Fermi National Accelerator Laboratory, Batavia,            
                   Illinois 60510}                                            
\centerline{$^{15}$Florida State University, Tallahassee, Florida 32306}      
\centerline{$^{16}$University of Hawaii, Honolulu, Hawaii 96822}              
\centerline{$^{17}$University of Illinois at Chicago, Chicago,                
                   Illinois 60607}                                            
\centerline{$^{18}$Indiana University, Bloomington, Indiana 47405}            
\centerline{$^{19}$Iowa State University, Ames, Iowa 50011}                   
\centerline{$^{20}$Korea University, Seoul, Korea}                            
\centerline{$^{21}$Kyungsung University, Pusan, Korea}                        
\centerline{$^{22}$Institute of Nuclear Physics, Krak\'ow, Poland}            
\centerline{$^{23}$Lawrence Berkeley National Laboratory and University of    
                   California, Berkeley, California 94720}                    
\centerline{$^{24}$Louisiana Tech University, Ruston, Louisiana 71272}        
\centerline{$^{25}$University of Maryland, College Park, Maryland 20742}      
\centerline{$^{26}$University of Michigan, Ann Arbor, Michigan 48109}         
\centerline{$^{27}$Michigan State University, East Lansing, Michigan 48824}   
\centerline{$^{28}$Institute for Theoretical and Experimental Physics,        
                   Moscow, Russia}                                            
\centerline{$^{29}$Moscow State University, Moscow, Russia}                   
\centerline{$^{30}$University of Nebraska, Lincoln, Nebraska 68588}           
\centerline{$^{31}$New York University, New York, New York 10003}             
\centerline{$^{32}$Northeastern University, Boston, Massachusetts 02115}      
\centerline{$^{33}$Northern Illinois University, DeKalb, Illinois 60115}      
\centerline{$^{34}$Northwestern University, Evanston, Illinois 60208}         
\centerline{$^{35}$University of Notre Dame, Notre Dame, Indiana 46556}       
\centerline{$^{36}$University of Oklahoma, Norman, Oklahoma 73019}            
\centerline{$^{37}$University of Panjab, Chandigarh 16-00-14, India}          
\centerline{$^{38}$Institute for High Energy Physics, Protvino 142284,        
                   Russia}                                                    
\centerline{$^{39}$Purdue University, West Lafayette, Indiana 47907}          
\centerline{$^{40}$Rice University, Houston, Texas 77005}                     
\centerline{$^{41}$Universidade do Estado do Rio de Janeiro, Brazil}          
\centerline{$^{42}$University of Rochester, Rochester, New York 14627}        
\centerline{$^{43}$CEA, DAPNIA/Service de Physique des Particules,            
                   CE-SACLAY, Gif-sur-Yvette, France}                         
\centerline{$^{44}$Seoul National University, Seoul, Korea}                   
\centerline{$^{45}$State University of New York, Stony Brook,                 
                   New York 11794}                                            
\centerline{$^{46}$Tata Institute of Fundamental Research,                    
                   Colaba, Mumbai 400005, India}                              
\centerline{$^{47}$University of Texas, Arlington, Texas 76019}               
\centerline{$^{48}$Texas A\&M University, College Station, Texas 77843}       
}                                                                             

\maketitle

\begin{abstract}

The shape of the transverse momentum distribution of $W$ bosons ($p_T^{W}$)
   produced in {\mbox{$p\bar p$}}\ collisions at {\mbox{$\sqrt{s}$ =\ 1.8\ TeV}}
   is measured with the D\O\ detector at Fermilab.
   The result is compared to QCD perturbative and resummation calculations over
   the $p_T^{W}$ range from $0-200\;\rm GeV/\it c$. The
   shape of the distribution is consistent with  the theoretical
   prediction.

\end{abstract}

\pacs{PACS numbers 12.38.-t, 06.20.Jr}

The transverse momentum ($p_T^{W}$) of $W$ intermediate vector bosons 
    produced in proton-antiproton collisions is due to the production of
    one or more gluons or quarks along with the boson.
    At low transverse momentum ($p_T^{W} < 10\;\rm GeV/\it c$), 
    multiple soft gluon emission is expected to dominate the 
    cross section. A soft gluon resummation technique~[1-5] is therefore
    used to make QCD predictions.  At high transverse momentum
    ($p_T^{W} > 20\;\rm GeV/\it c$), the cross section is dominated by the
    radiation of a single parton with large transverse momentum. 
    Perturbative QCD~\cite{ARtheory} is therefore expected to be reliable in
    this regime.  A prescription~\cite{AKtheory}  
    has been proposed for matching
    the low and high $p_T^{W}$ regions to provide a
    continuous prediction for all $p_T^{W}$.
    Thus, a measurement of the transverse momentum distribution may be used to
    check the soft gluon resummation calculations in the low $p_T^{W}$ 
    range, and to test the perturbative QCD calculations at high $p_T^{W}$.

The transverse momentum spectrum of $W$ bosons has been 
    measured previously by the UA1~\cite{UA1}, UA2~\cite{UA2},
    and CDF~\cite{CDF} collaborations, but with smaller data samples than
    the one used here.
    This paper presents a measurement of the shape of the $p_T$ 
    spectrum of $W$ bosons produced in {\mbox{$p\bar p$}}\ collisions at 
    {\mbox{$\sqrt{s}$ =\ 1.8\ TeV}} with the D\O\ detector~\cite{D0detector}
    at Fermilab, and extends the $p_T^{W}$ range of the previous measurements. 
    The data come from a sample of  $12.4\pm0.7\;\rm pb^{-1}$
    collected during the 1992--1993 run. A measurement of the inclusive cross
    section for $W$ and $Z$ boson production based on the same data set has been
    reported~\cite{xsecprl} and agrees with QCD predictions.

This measurement uses the decay mode $W\rightarrow e\nu$.
    Electrons were detected in a hermetic uranium/liquid-argon calorimeter 
    with an energy resolution of about $15\%/\sqrt{E(\rm GeV)}$.
    The calorimeter has a transverse granularity of $\Delta\eta \times 
    \Delta\phi = 0.1 \times 0.1$, where $\eta$ is the pseudorapidity 
    and $\phi$ is the azimuthal angle.
    Electrons were accepted in the central pseudorapidity region only,
    $|\eta|<1.1$, to keep the background contamination from multijet
    events at a reasonably low level for high values of $p_T^{W}$.
    The transverse momentum of the neutrino was calculated using
    the calorimetric measurement of the missing 
    transverse energy ({\mbox{$\not\!\!E_T$}}) in the event.  
    We take the $p_T^{W}$ to be the sum of the electron and neutrino transverse
    momenta, measuring it only from the recoiling hadrons.
    The analysis used a single electron trigger, 
    which required one electron with transverse
    energy ($E_T$) greater than $20\;\rm GeV$. 

The offline electron identification requirements consisted of four criteria: 
(i) the electron had to deposit at least $95\%$ of its
energy in the electromagnetic calorimeter (21 radiation lengths deep); (ii) the
transverse and longitudinal shower shapes had to be consistent with those
expected for an electron~\cite{topprd}; (iii) a good
    match had to exist between a reconstructed track in the drift
    chamber system and the shower position in the calorimeter;
and (iv)
the electron had to be isolated 
from other energy deposits in the calorimeter, with $I<0.1$. This isolation
variable is defined as $I=[E_{\rm TOT}(0.4)-E_{\rm EM}(0.2)]/E_{\rm EM}(0.2)$,
where $E_{\rm TOT}(0.4)$ is the total calorimeter energy inside a cone of radius
$\sqrt{\Delta\eta^2+\Delta\phi^2}=0.4$ and $E_{\rm EM}(0.2)$ 
is the electromagnetic energy inside a cone of radius 0.2.
    To select the $W$ boson candidate sample, we 
    required one electron with $E_T>25\;\rm GeV$, and {\mbox{$\not\!\!E_T$}}
    $>25\;\rm GeV$.   
    Events having a second  electron with $E_T > 20\;\rm GeV$ that satisfies
    criteria (i), (ii), and (iv) were excluded from the candidate sample as 
    possible {\mbox{$ Z\rightarrow {e^+e^-}$}}\ events. Criterion (iii) was not
    applied to this second electron in order to allow for possible tracking
    inefficiencies. These selection criteria yielded 7132 $W\rightarrow e\nu$
    candidates.

    The trigger and selection efficiencies were determined using
    {\mbox{$ Z\rightarrow {e^+e^-}$}}\ events in which one of the electrons 
    satisfied the trigger and selection criteria. The second electron then
    provided an unbiased sample with which to measure the efficiencies. 
    No dependence of the trigger or selection efficiency on 
    $p_T^{W}$ was found, to an accuracy of 5\%.

A Monte Carlo program~\cite{wmassprd} 
was used to simulate the D\O\ detector response and
   calculate the kinematic and geometric acceptance
   as a function of $p_T^{W}$. 
   The detector resolutions used in the
   Monte Carlo were determined from data, and were parametrized as  
   a function of energy and angle. The relative response of the hadronic and EM
calorimeters was studied using the transverse momentum of the $Z$ boson
as measured by the $p_T$ of the two electrons compared to the hadronic recoil 
system in the $Z$ event. 
This parametrized representation of the D\O\ detector was
used to smear the theoretical prediction by detector effects and 
compare it to our measured $p_T^{W}$. We prefer this method of comparison
to a standard unfolding procedure~\cite{dagostini} 
that proved to be sensitive to the
choice of the prior distribution function.  This sensitivity is caused by the
Jacobian zero in $dN/dp_T^W$ at the origin, which induces a peak that
appears near $p_T^W \approx 4\;\rm GeV/\it c$ 
after it is broadened by our $p_T^W$ resolution.  Only below $4\;\rm GeV/\it c$
do the true $p_T^W$ distributions predicted by 
the two available models~\cite{AKtheory,LYtheory}
exhibit a difference, which is masked in the
data by these same resolution effects.

The dominant source of background in the $W$  boson sample
    was multijet events where one or more of the jets fluctuated to
    fake an electron.  Some multijet events also have significant 
    {\mbox{$\not\!\!E_T$}} due to fluctuations and mismeasurements of the
    jet energies. This could fake a neutrino from $W$ boson decay.
    The amount of multijet background in the $W\rightarrow e\nu$ candidate
    sample was determined by first defining a ``loose'' event sample which
    had the same selection criteria as the candidate sample except that
    electron identification criteria (i) and (ii) were not applied.
This loose sample consisted of $N_{\rm s}$ signal events and $N_{\rm b}$ 
multijet background events. The $W\rightarrow e\nu$ candidate sample (described
above) consisted of  $\varepsilon_{\rm s} N_{\rm s}$ signal events and
$\varepsilon_{\rm b} N_{\rm b}$ multijet background events, where
$\varepsilon_{\rm s}$ and $\varepsilon_{\rm b}$ are the electron selection  
efficiencies for the candidate relative to the loose samples, for signal and 
background respectively. We obtained $\varepsilon_{\rm s}$  from the 
{\mbox{$ Z\rightarrow {e^+e^-}$}}\ sample, and 
$\varepsilon_{\rm b}$ from events with {\mbox{$\not\!\!E_T$}}
$<15\;\rm GeV$. 
    The total multijet background was determined to be $(4.2\pm2.3)\%$.
    The shape of the multijet background as a function of 
    $p_T^{W}$ was determined by repeating the procedure in different $p_T^{W}$ 
    bins.

\begin{figure}
\centerline{\psfig{bbllx=93pt,bblly=144pt,bburx=510pt,bbury=666pt,figure=
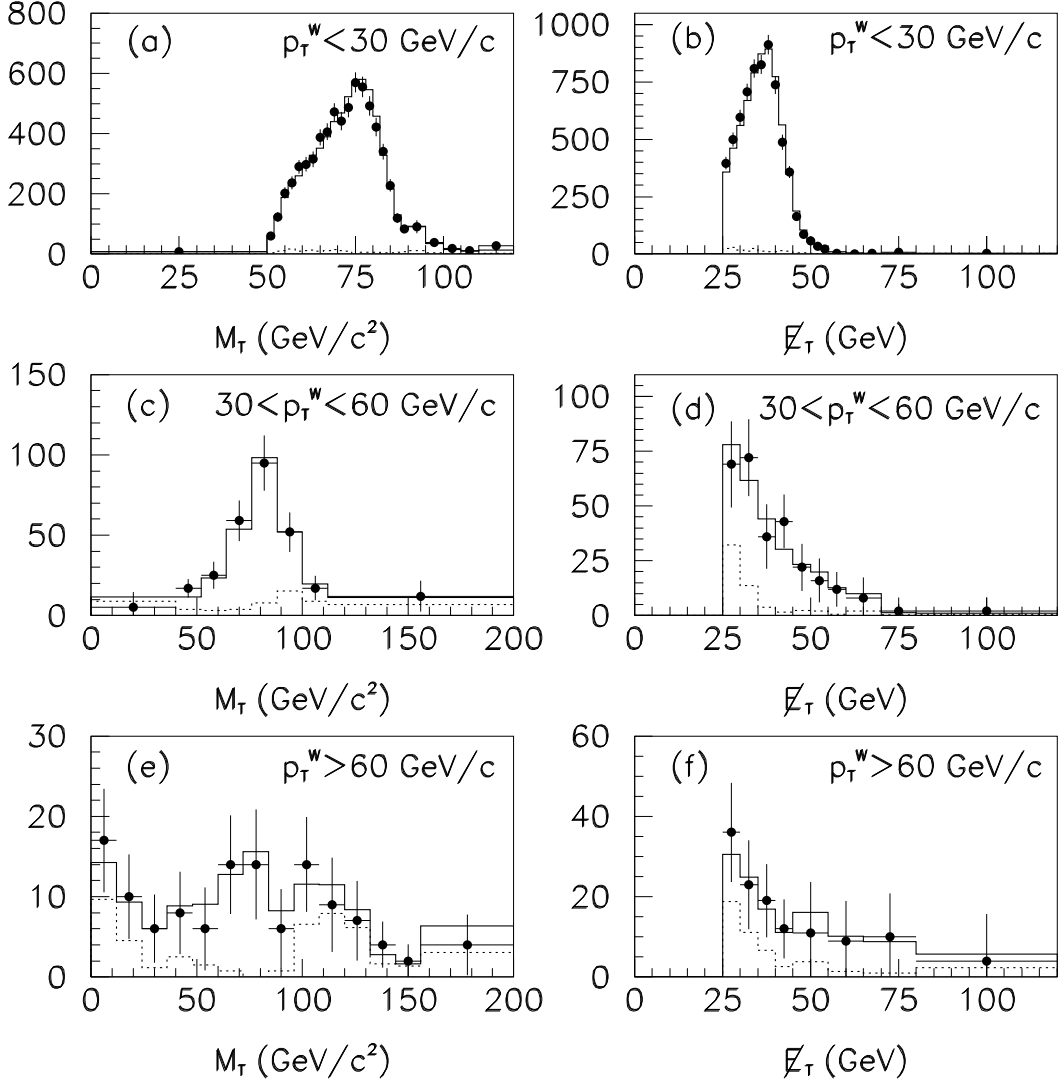,width=12.0cm}}
\caption{The transverse mass (left) and {\mbox{$\not\!\!E_T$}} (right) 
distributions for three $p_T^{W}$ bins. The points are the D\O\ data. The
solid histogram is the sum of the Monte Carlo signal and the estimated
background. The dotted histogram is the estimated background alone.}
\label{fig:fits}
\end{figure}

To cross--check the multijet background estimate, the
transverse mass and the {\mbox{$\not\!\!E_T$}} distributions of the final data
sample were compared to a model of the expected distributions
obtained from a combination of $W\rightarrow e\nu$ Monte Carlo events
plus the estimated multijet background.
The comparison was performed in three $p_T^{W}$ bins: $0-30\;\rm GeV/\it c$, 
$30-60\;\rm GeV/\it c$, and $>60\;\rm GeV/\it c$;
the number of $W\rightarrow e\nu$ candidates in each bin was
6726, 282, and 124.
The amount of multijet background in 
each $p_T^{W}$ bin was estimated as $(2.9 \pm 1.6)\%$, $(20.9 \pm 11.7)\%$, 
and $(38.3 \pm 21.5)\%$, respectively.
Figure~\ref{fig:fits} shows the results of the comparisons.
The distinctive shape of the transverse mass 
distribution for multijet background arises from applying the
kinematic cuts and the minimum $p_T^{W}$ requirement to a 
 sample predominantly composed of dijet events that lie
back--to--back in the transverse plane.
The goodness of the fit between the data and the model
is evaluated by performing a $\chi^2_{\lambda}$
test~\cite{chisql}. The numerical results 
for $\chi^2_{\lambda}/ \rm d.o.f.$ for each fit in Fig.~\ref{fig:fits} are
${\rm (a)}\; 20.2/25$, ${\rm (b)}\; 26.4/19$, ${\rm (c)}\; 4.7/7$, 
${\rm (d)}\; 5.4/9$, ${\rm (e)}\; 3.0/13$, and ${\rm (f)}\; 2.1/7$.  
They correspond to fit probabilities of
${\rm (a)}\; 74\%$, ${\rm (b)}\; 12\%$, ${\rm (c)}\; 69\%$,
${\rm (d)}\; 80\%$, ${\rm (e)}\; 99\%$, and ${\rm (f)}\; 95\%$
respectively. We therefore conclude that
the tests show good agreement between the data and the expectation. 

The normalized multijet background
    was subtracted bin by bin from the $W$ boson candidate sample transverse
    momentum spectrum.
    Additional corrections (all less than 5\%) were made to account for
    top quark background events and for 
    {\mbox{$ Z\rightarrow {e^+e^-}$}}\ events where one of the
    electrons was lost or not identified.
    Since $p_T^{W}$ was measured from the recoiling hadrons,
    the events originating from 
    $W\rightarrow \tau\nu$ (where $\tau\rightarrow e \nu\nu$) 
    contributed properly to the differential distribution; this source
of background therefore was included in the Monte Carlo simulation of the
$p_T^{W}$ distribution.

\begin{figure}[t]
\centerline{\psfig{bbllx=93pt,bblly=144pt,bburx=510pt,bbury=666pt,figure=
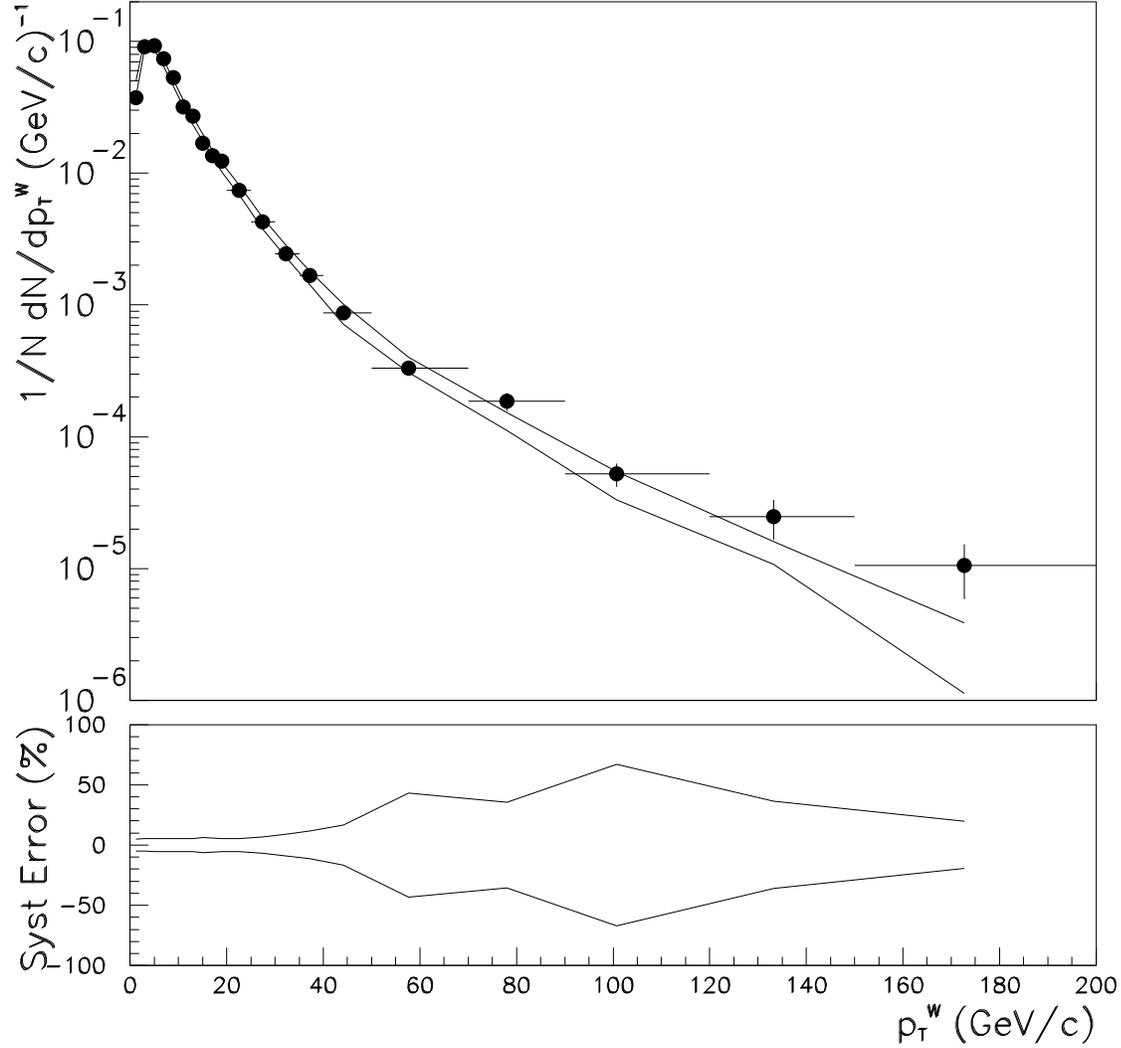,width=12.0cm}}
\caption{The $W$ boson transverse momentum spectrum, showing the
D\O\ result (solid points) with statistical uncertainty.
The theoretical calculation by Arnold and Kauffman~[4], smeared for detector
resolutions, is shown as two lines corresponding to the $\pm 1\sigma$ variations
of the uncertainties in the detector modelling. 
Within each bin, the values are plotted at the mean $p_T^{W}$.
The fractional systematic uncertainty on the data 
is shown as a band in the lower portion of the plot. The values of the
uncertainties for different $p_T^{W}$ bins are $100\%$ correlated with each
other. Upward fluctuations in the magnitude of the multijet background cause
the widening observed in the band at about 60 and $100\;\rm GeV/\it c$ in
$p_T^{W}$.}
\label{fig:ptw}
\end{figure}

\begin{figure}
\centerline{\psfig{bbllx=93pt,bblly=144pt,bburx=510pt,bbury=666pt,figure=
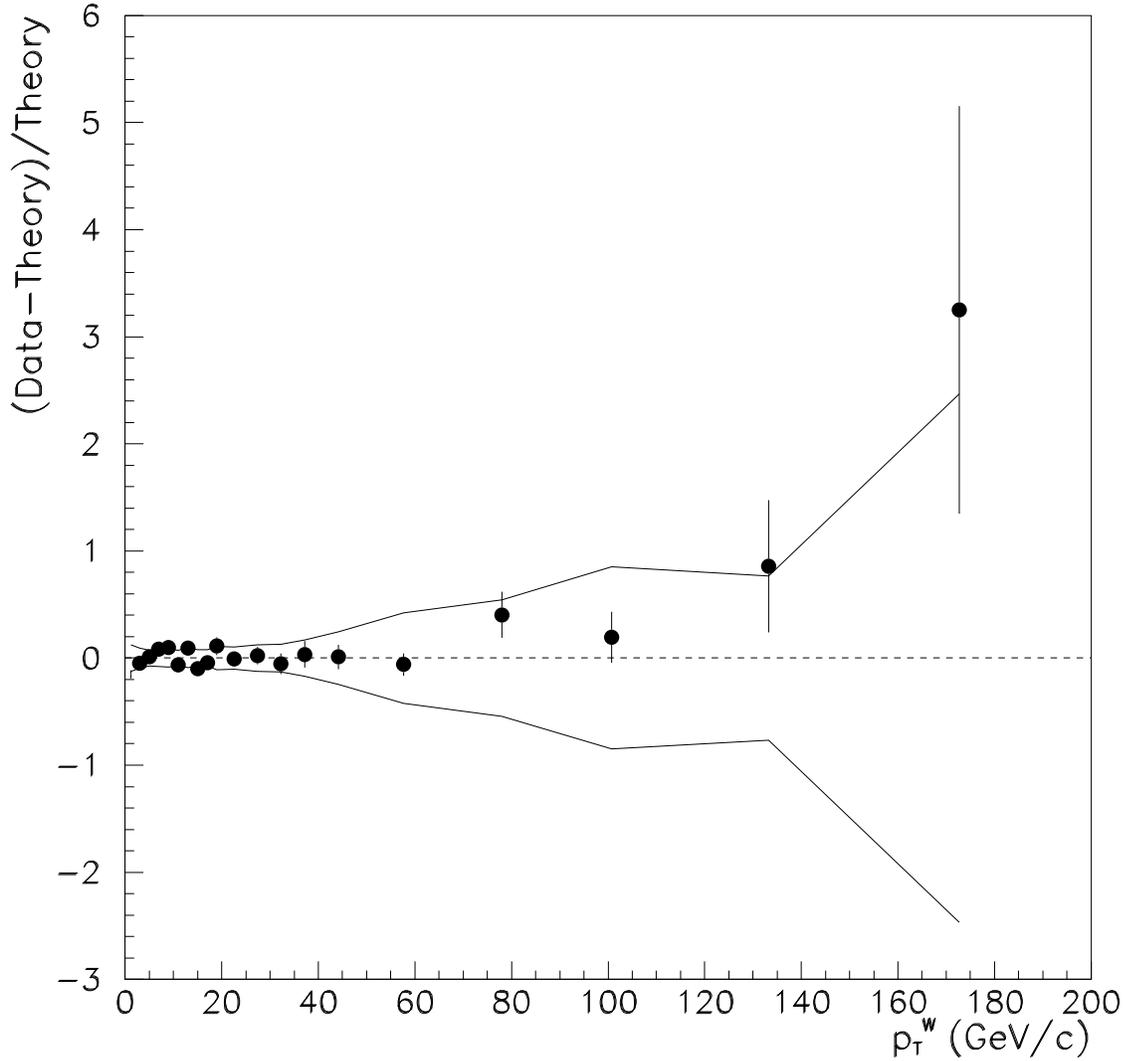,width=12.0cm}}
\caption{The ratio (Data--Theory)/Theory shown as a function of
$p_T^{W}$ with its statistical uncertainty as error bars. 
Within each bin, the values are plotted at the mean $p_T^{W}$.
The theory corresponds to reference~[4], smeared for detector
resolutions. The systematic uncertainties from data (background and efficiency)
and from the detector modelling are added in quadrature and shown 
as a band.}
\label{fig:ratio}
\end{figure}

The normalized distribution of the $W$ boson transverse momentum 
({\mbox{$\frac{1}{N}\frac{dN}{dp_T^{W}}$}}) is shown 
   in Fig.~\ref{fig:ptw} and given
   in Table~\ref{tab:ptw}. The largest contributions to the 
   systematic error in the $p_T^{W}$ measurement are the uncertainty
   in the magnitude of the multijet background, the uncertainty
   in the hadronic recoil energy scale factor and resolution used in the
   detector simulation, and the uncertainty in the selection efficiency. 
   These are all added in quadrature since they are independent. 
   We compare our experimental
   result to the theoretical prediction~\cite{AKtheory}
   computed using the MRSA$'$~\cite{mrsa} 
   parton distribution function and smeared
   for detector resolutions. The measurement and the prediction are 
   independently area--normalized to unity. These points are used to
   perform a more detailed comparison between data and
   theory by plotting the ratio (Data--Theory)/Theory, which
   is shown in Fig.~\ref{fig:ratio}. Data and theory differ by less than a half
   of a standard deviation above $p_T^W$ of $60\;\rm GeV/\it c$. 
   We therefore conclude
   that the shape of the distribution is consistent with  the theoretical
   prediction.

In summary, we have measured the shape of the transverse momentum distribution
of  
   $W$ bosons produced in {\mbox{$p\bar p$}}\ collisions at 
   {\mbox{$\sqrt{s}$ =\ 1.8\ TeV}}, and have found that it
   is consistent with the combined
   QCD perturbative and resummation calculations.  

%
We thank the staffs at Fermilab and collaborating institutions for their
contributions to this work, and acknowledge support from the 
Department of Energy and National Science Foundation (U.S.A.),  
Commissariat  \` a L'Energie Atomique (France), 
State Committee for Science and Technology and Ministry for Atomic 
   Energy (Russia),
CAPES and CNPq (Brazil),
Departments of Atomic Energy and Science and Education (India),
Colciencias (Colombia),
CONACyT (Mexico),
Ministry of Education and KOSEF (Korea),
and $\rm CONICET$ and UBACyT (Argentina).

\begin{table*}[t]
\caption{The $W$ boson transverse momentum  distribution
corresponding to Fig.~2. The column labeled ``Stat Error'' shows the statistical
uncertainty; ``Syst Error'' shows the
systematic uncertainty in background and efficiency;
``Detector Error'' shows the systematic uncertainty in
the detector modelling; ``Total Error'' is the sum in quadrature of the previous
three columns.}
\begin{center}
\begin{tabular}{c d d d d d d d}
Bin Width   & $\langle p_T^W \rangle$ & $N_{\rm signal}$ & 
{\mbox{$(1/N)(dN/dp_T^{W})$}} & Stat & Syst & Detector & Total \\ 
  &  &  & & Error & Error & Error & Error \\
\scriptsize{(GeV/{\it c})}        & \scriptsize{(GeV/{\it c})}        &     &  
\scriptsize{({\it c}/TeV)} & \scriptsize{({\it c}/TeV)} &  
\scriptsize{({\it c}/TeV)} & \scriptsize{({\it c}/TeV)} &
\scriptsize{({\it c}/TeV)} \\  \hline 
2& 1.2 &  506.8 &  37.4 & 1.6 & 1.9 & 6.1 & 6.6 \\ 
2& 3.0 & 1232.0 &  90.8 &  2.6 & 4.6 & 7.9 & 9.5 \\ 
2& 5.0 & 1253.0 & 92.4 & 2.6 & 4.8 & 4.4 & 7.0  \\ 
2& 7.0 & 1006.6 & 74.2 &  2.3 & 3.9 & 3.7 & 5.8 \\ 
2& 9.0 &  718.4 & 53.0 &   1.9 & 2.8 & 2.6 & 4.2 \\
2& 11.0 &  431.2 & 31.8 &   1.5 & 1.7 & 2.0 & 3.0 \\
2& 13.0 &  368.4 & 27.2 &  1.4 & 1.4 & 1.6 & 2.6  \\
2& 15.1 &  228.0 & 16.8 &  1.1 & 1.1 & 1.2 & 1.9 \\ 
2& 17.1 &  184.4 & 13.6 &   1.0 & 0.8 & 0.8 & 1.5 \\
2& 19.0 &  167.9 & 12.4 &  0.9 & 0.7 & 0.9 & 1.5 \\ 
5& 22.6 &  252.0 &  7.43 & 0.46 & 0.42 & 0.65 & 0.89 \\
5& 27.3 &  145.4 & 4.29 &   0.34 &  0.30 & 0.43 & 0.61 \\
5& 32.3 &  83.0 & 2.45 &   0.25 & 0.22 & 0.26 & 0.42 \\ 
5& 37.2 &  56.7 & 1.67 &  0.20 &  0.19 & 0.18 & 0.33 \\ 
10& 44.2 & 59.4 & 0.875 & 0.099 & 0.147 & 0.150 & 0.232 \\
20& 57.7 & 45.2 & 0.333 & 0.037 & 0.144 & 0.045 & 0.155 \\
20 & 78.0& 25.2 & 0.186 & 0.028 &0.066 & 0.020 & 0.075  \\
30 & 100.7 & 10.7 &  0.052 & 0.011 & 0.035 & 0.010 & 0.038 \\
30 & 133.2 &  5.1 & 0.025 & 0.008 & 0.009 & 0.002 & 0.013 \\
50 & 172.7 & 3.6 & 0.011 &  0.005 & 0.002 & 0.001 & 0.005 \\    
\end{tabular}
\end{center}
\label{tab:ptw}
\end{table*}

\clearpage

\end{document}